\documentclass[12pt,a4paper]{article}
\usepackage{amsmath,amssymb}
\usepackage{graphics}
\usepackage{epsfig}
\usepackage{graphicx}
\newcommand{\be}{\begin{equation}}
\newcommand{\ee}{\end{equation}}
\newcommand{\bea}{\begin{eqnarray}}
\newcommand{\eea}{\end{eqnarray}}
\newcommand{\beas}{\begin{eqnarray*}}
\newcommand{\eeas}{\end{eqnarray*}}
\newcommand{\ba}{\begin{array}}
\newcommand{\ea}{\end{array}}
\newcommand{\nn}{\nonumber}

\newcommand{\bt}{\begin{table}}

\newcommand{\al}{\alpha}
\newcommand{\ga}{\gamma}

\newcommand{\Ga}{\Gamma}	

\newcommand{\de}{\delta}
\newcommand{\De}{\Delta}

\newcommand{\ka}{\kappa}
\newcommand{\la}{\lambda}
\newcommand{\La}{\Lambda}
\newcommand{\na}{\nabla}
\newcommand{\om}{\omega}
\newcommand{\Om}{\Omega}

\newcommand{\si}{\sigma}

\newcommand{\g}{\sqrt{-g}}

\begin{document}
\title{
{\bf 
Quartet-metric/multi-component gravity: 
scalar graviton as emergent  dark  substance
}}
\author{Yury~F.~Pirogov
%\footnote{E-mail: yury.pirogov@ihep.ru }
\\
{\small
\em 
Theory Division, 
Institute for High Energy Physics of
NRC Kurchatov Institute,}\\
{\small
\em  Protvino 142281, Russia 
}
%Moscow Region,
%\\ \small{\em E-mail:  yury.pirogov@ihep.ru}
}
\date{}
\maketitle
\begin{abstract}
\noindent
In the general frameworks of an earlier introduced 
quartet-metric/multi-component gravity, a  theory 
of a massive scalar  graviton supplementing 
the  massless tensor one is consistently deduced.
The peculiarities of the scalar-graviton field 
compared to the canonical scalar one are demonstrated.
The  (ultra-)light  scalar graviton is treated  
as an emergent  dark substance of the Universe: 
dark matter and/or dark energy 
depending on the solution. The case with 
scalar graviton as dark energy  
responsible for  the late-time accelerated expansion 
of the Universe is studied in more detail.
In particular,  it is shown that due to 
an attractor solution for the light scalar graviton
there naturally emerges  at the classical level  a tiny  
nonzero effective cosmological constant, 
even in the absence of the Lagrangian one. 
The prospects of going beyond LCDM model per scalar graviton 
are shortly indicated.\\


%{\bf PACS:} {{04.50.Kd} {Modified theories of gravity}; % \and 
%{95.35.+d} { %\and
%{95.36.+x}  { Dark energy}
%} % end of PACS codes
%} %end of abstract

\end{abstract}

\section{Introduction}
\label{intro}

The present-day  scenario for the evolution  of the Universe is described  by
the so-called Cosmological  Standard  Model  or,
more particularly, the $\La$CDM model. The latter incorporates, in accord with
its name, such  new ingredients  as the cosmological constant (CC)
$\La$ and a  (cold) dark matter (DM) of an unknown nature.  Being extremely economic
in  the theoretical concepts the model shows an impressive
phenomenological success in describing the observational data.
Nevertheless, some arguments (though mainly of the the  theoretical nature) concerning 
CC may imply the necessity of  going eventually 
beyond such a  model, in particular, through  a
hypothetical  dark  energy (DE) superseding CC.\footnote{For $\La$CDM 
and beyond, see, e.g.,~\cite{Joyce, Bull}.}  
With the $\La$CDM  model being 
 based on General Relativity  (GR) as a working tool,
going beyond the model  may,  in particular,  imply going beyond
GR\footnote{For the modified and extended theories  of gravity
beyond GR, see, e.g.,~\cite{Capo}--\cite{Calm}.} 
in looking for explanation of DM and DE.

In this vein, in refs.~\cite{Pir1,Pir2}  there was proposed  
an  effective field theory (EFT) of 
the {\em quartet-metric/multi-component  gravity}\/.
The theory  is based on the two physical concepts. ({\em i}\/)
There exist in spacetime some  distinct dynamical 
coordinates, given by a scalar quartet,  
playing the role of the Higgs-like fields for gravity. 
The number of the original gravitational components 
increases thus to fourteen: ten for metric and four for scalars 
(in the four spacetime dimensions).
({\em ii}\/) The diffeomorphism invariance of the quartet-metric GR 
gets (partially) spontaneously broken/hidden, 
with the gauge components contained in  metric 
becoming physical through absorbing the scalar quartet. 
In a general case, such a multi-component theory 
describes in a  fully dynamical and    
generally-covariant (GC) fashion 
the (massive) scalar-vector-tensor  gravity,
with the additional physical gravity 
components a priori  serving  as DM and/or DE  (and, conceivably, beyond)
depending on the solution.
It was argued that the mere admixture to metric of the scalar quartet
may  result in the wide  variety of the particular versions of the theory,  
with an extremely  rich spectrum of the emergent 
physical phenomena beyond GR.  To systematically 
study the latter ones,  with the aim of picking-out 
the most  relevant version of the theory (if any),  presents  a big  challenge.
Though, in the spirit of the so-called  Occam's razor,   
the most likely version of  the multi-component gravity to serve as the next-to-GR one
may be   given  by that  just with the massive scalar graviton  
supplementing the conventional  (massless) tensor one of GR.
So, in the present paper we  systematically  adhere  
to such a line of reasoning.\footnote{Of course, this by no means 
deprives  other GR modifications  conceivable within  the multi-component gravity, 
such as, say,  the pure-tensor gravity with  the massive 
(tensor) graviton possessing, in particular,  a modified kinetic term, etc,~\cite{Pir1, Pir2}.}

 In section~2, there is presented  the genesis of a consistent 
scalar-graviton theory  staring from the first principles of the more general   
multi-component gravity.
This allows to clarify the nature of the (otherwise ad hoc)
scalar graviton, as well as  to open prospects 
for its possible future modifications and generalizations.
To this end,   the multi-component gravity is concisely exposed, with   
its scalar-graviton reduction consistently deduced. 
The relation  of the latter with and distinction 
from the Horndeski  scalar-tensor theory~\cite{Berg, Horn} 
is exposed. In section~3, 
the scalar graviton is  worked-out  as a dark substance,  with a set  of 
simplifications step-by-step imposed, 
and the  peculiarities of the scalar graviton  
compared to a canonical scalar field are demonstrated. In particular, 
the attributes of the scalar-graviton 
solutions required for the scalar graviton  
to serve as  DM or DE are shown.
In section~4, the scalar-graviton field   is applied as DE 
filling-up homogeneously  the whole 
Universe on the late-time stage of its expansion. 
In particular,  a mechanism of producing the tiny  nonzero effective CC 
through an attractor solution for the (ultra-)light scalar graviton is put forward.
In Conclusion, the necessity of further studying the scalar graviton 
to validate it is as an emergent dark substance of the Universe 
is stressed.

\section{Multi-component gravity and scalar graviton}
\label{sec:2}

\subsection{Multi-component gravity: generalities}

Let us  start with a concise  exposition of 
EFT of the quartet-metric/multi-component 
gravity~\cite{Pir1, Pir2}, the latter  reducing   in a limit to the metric  GR. 
Such a theory  is generically given by  a  GC action
\be\label{S}
I = \int {\cal L}_G(g_{\mu\nu}, g, \partial_\mu \om^a, \eta_{ab}) \, d^4 x ,
\ee
with  a  Lagrangian scalar density ${\cal  L}_G$ for  the extended gravity 
dependent on the metric $g_{\mu\nu}$ ($g\equiv \det (g_{\mu\nu})<0$) 
and  a quartet of the  scalar fields  
${\om}^{a}$, $a=0,\dots,3$.
At that,   $a,b,\dots$ are the  indices of the global   Lorentz
symmetry  $SO(1,3)$,  with the  invariant  Minkowski symbol 
$\eta_{ab}$.  By default, the
signatures of $g_{\mu\nu}$ and $\eta_{ab}$ are  chosen  to  coincide.  The
scalar fields ${\om}^{a}$ are defined up to the (patch-wise) 
global Poincare transformations 
(independent of the spacetime)
composed of the Lorentz
ones  and shifts ${\om}^{a}\to {\om}^{a}+c^{a}$, with the  arbitrary
constant parameters~$c^{a}$.
The quartet $\om^a$ defines  the  (patch-wise)  
invertible coordinate  transformations in spacetime 
from the arbitrary observer's coordinates $x^\mu$ to  some  distinct 
dynamical world coordinates $\mathring{x}^a$, the so-called,   {\em quasi-affine} ones:
 $\mathring{x}^{a}={\om}^{a}(x)$   
 (with  the inverse $x^\mu=x^\mu(\mathring{x})$).   
Physically,  such  coordinates may 
 be considered   as comoving
with the vacuum treated ultimately as a dynamical system  on par 
with the observable world. 
Due to GC and the global Poincare invariance, $\om^a$ enters, in fact, 
 through an auxiliary  {\em quasi-Lorentz} metric  
\be
{\om}_{\mu\nu}\equiv\partial_\mu  {\om}^a  \partial_\nu {\om}^b \eta_{ab},
\ee
with
\be\label{key}
{\om}\equiv \det(\om_{\mu\nu})  = \det (\partial_\mu 
{\om}^a)^2 \det(\eta_{ab})<0.
\ee
To ensure the  (patch-wise) invertibility 
of the spacetime coordinate transformations 
$\mathring{x}^a=\om^a(x)$ one should have the Jacobian $\det (\partial_\mu 
{\om}^a)\neq 0$ and thus ${\om}\neq 0$, implying the  non-degeneracy of the quasi-Lorentz 
metric $\om_{\mu\nu}$, with  an inverse 
$\om^{-1\mu\nu}$.\footnote{Not to mix $\om^{-1\mu\nu}$ with 
$\om^{\mu\nu}\equiv g^{\mu\ka} g^{\nu\la} \om_{\ka\la} $.} 
In view of $\om\neq 0$, the sign of $\sqrt{-\om}$ is preserved and we can choose 
$\sqrt{-\om}>0$.
In these terms,  the Lagrangian  of  the multi-component gravity
may most generally be rewritten  in an entirely  spacetime  form~as

\be
{\cal L}_G  = {\cal L}_{G}(g_{\mu\nu}, 
\om_{\mu\nu}, g, {\om}).
\ee
In particular, the kinetic terms beyond GR enter through the GC tensor given by 
the difference of the 
two Christoffel connections: 
\be
B^\la_{\mu\nu}=\Ga^\la_{\mu\nu}(g_{\mu\nu})- 
\ga^\la_{\mu\nu}(\om_{\mu\nu}),
\ee
including, in particular, 
\be
B^\la_{\mu\la}=\partial_\mu 
\ln \sqrt{-g}/\sqrt{-\omega}
\ee
for the  kinetic term of the scalar graviton.
The potential terms enter through the GC scalars built from the powers 
of the GC tensor 
\be
\Om_\mu{}^\nu=\om_{\mu\la} g^{\la\nu}
\ee
and its inverse $\Om^{-1}{}_\nu{}^\mu=g_{\nu\ka} \om^{-1\ka\mu}$,
as well as $\det(\Om_\mu{}^\nu)=\om/g$.
At that, the true dynamical variables remain still the metric $g_{\mu\nu}$ 
and the scalar quartet~$\om^a$. Ultimately, the latter serves as a Higgs-like field for 
gravity (partially) breaking the diffeomorphism symmetry 
of the quartet-metric/multi-component  gravity 
and making the gauge components of $g_{\mu\nu}$ physical.
Technically, the  quasi-affine coordinates $\mathring{x}^a$   
are distinct by the fact that under  using them 
the quasi-Lorentz metric gets  Minkowskian form,  
$\om_{ab}(\mathring{x})\equiv \eta_{ab}$ 
(respectively, $\om^{-1 ab}(\mathring{x})\equiv \eta^{ab}$). 
As a result,  the  auxiliary affine connection 
${\ga}^\la_{\mu\nu}$ becomes in these coordinates zero,
${\ga}^{c}_{ab}(\mathring{x})=0$. Yet, the 
Christoffel  connection  corresponding to the 
spacetime metric $g_{\mu\nu}$ in these coordinates,  
$\Ga^{c}_{ab}(\mathring{x})$,  remains, generally, nonzero. 

The Lagrangian density ${\cal L}_G$ may further be decomposed 
as 
\be
{\cal L}_G= L_G(g_{\mu\nu},  \om_{\mu\nu},g/{\om} ) \, {\cal M}(g, {\om}),
\ee
with $L_G $ a GC scalar  Lagrangian 
and ${\cal M}$ a GC scalar density 
 of the proper weight 
(a spacetime measure entering the spacetime volume element 
$d V= {\cal M}d^4 x$). 
The measure may generally be chosen as 
\be
{\cal M}=\varphi (g/ {\om})\sqrt {-g}
\equiv \varphi (g/ {\om})(g/{\om})^{1/2} \sqrt{-{\om}}, 
\ee
with $\varphi (g/ {\om})$ an arbitrary  function of the scalar $g/\om.$ 
With the proper  redefinition of $L_G$, the measure may   
equivalently  be chosen either as $\sqrt {-g}$ or $\sqrt {-{\om}}$, 
depending on the context.
So, prior fixing the Lagrangian we can  without loss of generality put 
\be
I= \int { L}_G(g_{\mu\nu},  \om_{\mu\nu},g/{\om} ) \sqrt{-g}\, d^4 x ,
\ee

An  $ L_G$ quadratic in the first
derivatives  of metric is considered  in~\cite{Pir1}.
Generally, such an  $ L_G$  describes the  massive scalar  ($s$),  vector  ($v$) 
and tensor ($g$) gravitons contained 
in the metric, with $\om^{a}$ serving ultimately as a
 gravity counterpart  of the Higgs fields.  
Imposing  on the  parameters of $L_{G}$ some ``natural'' (in a
technical sense) restrictions, one can exclude  in the
linear approximation   the vector
graviton as the most  ``suspicious''  theoretically and phenomenologically,  
leaving   in  this approximation in addition 
to the tensor graviton  just  the  massive scalar
one as the most ``auspicious''.
A more  general  multi-component gravity Lagrangian  is discussed in~\cite{Pir2}.
Such a Lagrangian admits the two  generic types of reductions significantly simplifying  
the theory: the scalar-graviton reduction and the second-derivative reduction.
To result in  as simple as possible version of the theory  we impose step-by-step   
both  types of reduction.

\subsection{Scalar-graviton reduction}   

\subsubsection{High-order  derivatives}  

The multi-component gravity is significantly simplified (remaining still rather
rich of the new content)
 under  considering  a reduced case   given by  the 
Lagrangian $L_G$ dependent on $\om_{\mu\nu}$
exclusively  through its determinant $\om$.
In fact, due to GC $\om$ should enter the Lagrangian   through  
the ratio  $\om/g $. 
Without loss of generality this ratio may be substituted by
\be\label{si}
\si\equiv \ln \sqrt{-g}/\sqrt{-{\om}}. 
\ee 
With $\om $  having the
same weight  as 
$g$ under the general coordinate  transformations,  
$\si$ is a true GC scalar field normalized to zero 
in a flat limit ($g={\om}=-1$). Stress that the scalar 
graviton $\si$ has a combined nature, ultimately distinguishing $\si$  
from an elementary scalar field,

In the scalar-tensor gravity with a canonical 
scalar field,  the most general Lagrangian of  the  arbitrary derivative order
in the metric and scalar field in the four spacetime dimensions, resulting still 
in the second-order  field equations (FEs)
in both fields is given in~\cite{Berg, Horn}.
The most general Lagrangian with the derivable  FEs which are
{\em quasi-linear} in the second derivatives of both the metric and scalar field (in
the sense that coefficients of the second derivatives contain no
derivatives) is developed by Horndeski~\cite{Horn}.\footnote{This 
is to avoid potentially possible 
Ostrogradsky instabilities for the higher then  second-order classical FEs,  
resulting in a theory with a ghost vacuum.}$^,$\footnote{At that, all the terms present 
in the Horndeski theory may be  shown to
originate  from the  scalar-tensor terms having Galilean symmetry in
the flat spacetime~\cite{Nic}. For a generalized Horhdeski's  
theory, see,~\cite{Charm0}.}

The respective Lagrangian for the scalar-reduced  multi-component gravity 
\be\label{Horn}
  L_{sg} =  L_{sg}(g_{\mu\nu},\si) ,
\ee
may be obtained   from the Horndeski one   by imposing the restriction~(\ref{si})
on the scalar field  or
by adding to (\ref{Horn}) a constraint  Lagrangian, say, in the form
\be\label{la}
L_\la=\la(e^{-\si}-\sqrt{-{\om}}/\sqrt{-g}), 
\ee
with $\la$ an indefinite Lagrange multiplier. This is a sole 
but crucial difference compared to  the original 
Horndeski scalar-tensor  theory resulting
in all the specifics of the scalar graviton.
Eqs.~(\ref{key}) and (\ref{si}), with
$g_{\mu\nu}$ and ${\om}^{a}$ as the  independent gravity 
field variables,  are the  key ingredients of the dynamical  theory of the scalar
graviton. For completeness, the so obtained pure-gravity Lagrangian
$L_{sg}$ is to be supplemented by
a matter one $L_m$ dependent on some  generic   matter
fields   $\phi^I$ and, generally,  on $\si$, too.
The  scalar-reduced multi-component gravity in terms of  
the Lagrangian~(\ref{Horn})
under constraint~(\ref{si}) or (\ref{la}) 
may be  proposed as the  next-to-GR EFT of gravity 
describing the  massless tensor graviton
supplemented by the massive scalar one. The latter  is assumed to 
serve  as an emergent  gravitational dark substance, in particular,  DM
and/or~DE depending on the solution.

\subsubsection{Second-order derivatives }  

Imposing additionally the second-derivative restriction 
we get  the tensor-scalar gravity Lagrangian as
\be\label{L}
L=L_{sg} +L_m =L_g(\partial_\la g_{\mu\nu},  g_{\mu\nu}, \si)
+L_s(\partial_\la \si, g_{\mu\nu},  \si)
+L_m(\partial_\la \phi^I,   \phi^I,  g_{\mu\nu},
\si).
\ee
More particularly, we put  in the second order 
\be\label{sg}
{\cal L}_{sg}=\Big[-\frac{1}{2}\ka_g^2 \varphi_g(\si)  R(g_{\mu\nu}) 
+\frac{1}{2} \ka_s^2\varphi_s(\si) 
g^{\mu\nu} \partial_{\mu}\si\partial_{\nu}\si -V_s (\si)\Big] \sqrt{-g},
\ee
were  $R(g_{\mu\nu})$ is the Ricci scalar,  
$\varphi_g> 0$ and $\varphi_s>0 $ are some arbitrary scalar functions,
and  $V_s$ is a scalar-graviton  potential, generally,
including as a constant part  a cosmological constant.
The parameters  $\ka_g $ and $\ka_s$ of 
the dimension of mass characterize  the strength, respectively, of 
the tensor and scalar gravity, with  $\ka_g=\ka_P\equiv 1/\sqrt{8\pi G_N}$  
given by  the reduced Planck mass.  
For the dominance  of tensor gravity  it is moreover assumed  that 
$\ka_s\ll \ka_g$. 
The function $\varphi_g$ characterizes the type of modification of the  tensor GR,
while $\varphi_s$, in fact, corresponds  to a redefinition  of the scalar graviton 
compared to (\ref{si}), the latter being  taken by  default as
appearing  naturally in  the multi-component gravity. 

Introducing the conformally rescaled  metric 
$\hat g_{\mu\nu}$ through 
\be
\hat  g_{\mu\nu}\equiv \varphi_g  (\si) g_{\mu\nu},
\ee
with $(-\det (\hat g_{\mu\nu}))^{1/2}  \equiv\sqrt{-\hat g}=\varphi_g^2 \sqrt{-g}$,
we can present (\ref{sg}) equivalently as 
\be\label{scal}
{\cal L}_{sg}=\Big[-\frac{1}{2}\ka_g^2   R(\hat g_{\mu\nu}) 
+\frac{1}{2} \ka_s^2\hat \varphi_s(\si) 
\hat g^{-1\mu\nu} \partial_{\mu}\si\partial_{\nu}\si -\hat V_s (\si)\Big] \sqrt{-\hat g},
\ee
with $ \hat\varphi_s$ and $\hat V_s$ properly redefined, 
and $\hat g^{-1\mu\nu}$ being an inverse of $\hat g_{\mu\nu}$.
In particular, putting $\varphi_g=e^{-\si/2}$ 
we get 
\be
\hat g_{\mu\nu}=  ({\om}/g)^{1/4} g_{\mu\nu},
\ee
implying $ \hat g= {\om}$.  This  case, supplemented 
by a properly modified matter Lagrangian, 
$\hat L_m(\phi^I, \hat g_{\mu\nu}, \si$), may be referred to
as  the quasi-\/{\em Weyl  transverse  gravity (WTDiff)}.\footnote{
For WTDiff, see, e.g.,~\cite{Iza}--\cite{Oda}.   Stress 
that   in  distinction with WTDiff,  $\om$ in qWTDiff 
is a dynamical variable,  $\om=\om(\om^a)$.
Clearly, under neglecting by the explicit dependence on $\si$,  qWTDiff 
reduces to GR with the redefined metric $\hat g_{\mu\nu}$. 
For matching  with   WTDiff (supplemented   
by a scalar graviton),   $\om$  in   
qWTDiff should  effectively be ``frozen''
to an auxiliary  non-dynamical/``absolute''  scalar density
by dropping-off the variation of the action with respect to $\om^a$. 
This, in fact,  means  abandoning a proper FE (cf., section 3).}

As a paradigm, we choose in what follows $\varphi_g=\varphi_s=1$
considering  the scalar-graviton modification of GR with the canonical $\si$.
Restricting ourselves by  $L_s$  at energies less than
$\ka_s$, we retain only the leading term in the
derivatives of~$\si$.
On the other hand, the scalar potential $V_s$ is still allowed to be an
arbitrary function of~$\si$.
Put $V_s(\si)  \equiv V_s|_{\rm min}+\De V_s(\si)$, 
where $V_s|_{\rm
min}\equiv \ka_g^2 \La$, with   $\La\ge 0$ being   a counterpart of the 
cosmological constant,  and $\De
V_s\ge0$. Under  $\De V_s\equiv 0$ the Lagrangian (\ref{scal}) 
gets at $\varphi_g=\varphi_s=1$
moreover  invariant under the global shifts $\si\to \si+ \si_0$, with arbitrary
constant $\si_0$. Thus, $\De V_s= 0$, as extending the symmetry of the Lagrangian,  
is natural in a technical sense, justifying the   relative
lightness of the scalar graviton. But this does not concern the
constant part  $ V_s|_{\rm min}$ which requires additional arguments 
in the favor of its absence/smallness.

\section{Scalar graviton as dark substance}

\subsection{General case}

Varying the Lagrangian density  with respect to 
$g_{\mu\nu}$, $\om^a$ and the generic matter fields $\phi^I$, 
and using, in particular,  the relations
\bea 
\de \si&=&  \de\sqrt{- g}/\sqrt{- g}  -   
\de\sqrt{-\om}/\sqrt{-\om},\nn\\
\de\sqrt{- g}&=&- (1/2)\sqrt{-g}\,{g}_{\ka\la}\de {{g}}^{\ka\la},\nn\\
\de\sqrt{- \om}&=& (1/2)\sqrt{-\om}\, \om^{-1\ka\la}\de
{{\om}}_{\ka\la},\nn\\
\de   {\om}_{\ka\la}  &=&  \eta_{ab}(\om^a_\ka \de  {\om}^b_\la
+{\om}^{a}_\la \de {\om}^b_\ka), 
\eea
where
$ {\om}^{-1\ka\la}={\om}^{-1}{}^\ka_{a}{\om}^{-1}{}^\la_{b} \eta^{ab} $ 
is an inverse of ${\om}_{\ka\la}$,  and ${\om}^{-1}{}^\la_{a}=
\partial x^\la/\partial \mathring{x}^a $ is
a tetrad\footnote{Not to mix ${\om}^{-1}{}^\la_{a}$
with  ${\om}^\la_{a}\equiv g^{\la\ka}\eta_{ab} {\om}_\ka^{b}$.}
inverse of  $ {\om}_\la^{a}\equiv\partial_\la \om^a$,
we   get  the system of 
the  coupled  FEs  for the metric, scalar quartet and
matter fields 
 in the conventional notation, respectively, as  
\bea\label{FEs}
G_{\mu\nu}\equiv R_{\mu\nu} 
-\frac{1}{2}R g_{\mu\nu}&=& \frac{1}{\ka_g^2}
(T_{s\mu\nu}  + T_{m\mu\nu} ) ,\nn\\
   \frac{\de }{\de {\om}^a}  (L_g+L_{s} +L_m)  & \equiv&  
\na_\la\Big(( \frac{   \de
L_s}{\de \si} +  \frac{\de L_m}{\de \si}  )
{\om}^{-1}{}^\la_{a}\Big)=0,\nn\\
  \frac{\de  L_m}{\de\phi^I}  &\equiv& \frac{\partial L_m}{\partial \phi^I}
-\na^\ka\frac{ \partial L_m}{\partial
\na_\ka\phi^I}=0,
\eea
where $\de/\de$ is a total variational derivative and $\partial/\partial$
a partial one. 
The first and the last  FEs  in (\ref{FEs}) are clearly the
counterparts of
the tensor gravity and matter FEs in GR,
The second FE for $\om^{a}$  is a generalization of  the scalar-field one. This follows 
from the fact that this
equation   embodies, in particular,   the ordinary one  for $\si$ in a
marginal case with  $\de(L_s+L_m )/\de \si=0$. 
But the latter should not fulfill 
in a general case. The specific   form of
the second FE, which looks like a  continuity  condition, 
is due to $L_G$  being dependent
only on the derivatives of ${\om}^{a}$ and thus invariant under the  global
shifts ${\om}^{a}\to {\om}^{a}  +c^{a}$, with any constant $c^{a}$. 
In the end, this implies a GC conserved current density  
${\cal J}^\mu_a=\sqrt{-g}(\de  L_s/\de\si  +\de  L_m/\de\si)  \om^{-1}{}^\mu_a$. 

The r.h.s.\ of the first of FEs (\ref{FEs}) may be treated as 
the total energy-momentum  tensor, with  $T_{s\mu\nu}$ and $T_{m\mu\nu}$   the
canonical 
energy-momen\-tum tensors, respectively, for the scalar graviton, 
as a kind of dark substance,   and the 
matter obtained by means of varying  the  Lagrangian $L_f$ of the respective   fraction 
$f=(s,m)$ through $g_{\mu\nu}$ as follows:
\be
T_{f\mu\nu} \equiv   \frac{2}{\g}\frac{\de(\g L_f)}{\de
g^{\mu\nu}}.
\ee
By this token, we get
\bea
T_{s\mu\nu}  &= &  \ka_s^2\na_\mu \si \na_\nu \si - 
\frac {1}{2}\ka_s^2
\na^\la\si \na_\la\si  g_{\mu\nu}  + U_s  g_{\mu\nu},\nn\\
T_{m\mu\nu}&=&
2\frac{\partial L_m}{\partial g^{\mu\nu}}-\Big( L_m +\frac{\de
L_m}{\de \si}\Big) g_{\mu\nu},
\eea
where $U_s$ is  a generalized potential 
\be
U_s\equiv V_s+W_s,
\ee
with
\be\label{W}
W_s\equiv- \de L_s/\de \si= 
\ka_s^2 \na^\la\na_\la \si +V'_s 
\ee
being the  rescaled  wave operator for the scalar field $\ka_s\si$, and the 
prime-sign meaning the derivative with respect to $\si$.
The reduced Bianchi identity, $\na_\nu
G^\nu_\mu=0$, results~in   
\be\label{Bian} \na_\mu T^\mu_\nu\equiv\na_\mu(T_s{}_{\nu}^\mu  
+ T_m{}_{\nu}^\mu ) =0
\ee
representing the covariant conservation/continuity of the total
energy-momentum tensor 
of matter supplemented by  the scalar graviton.
More particularly, (\ref{Bian}) proves to read 
\be\label{Bianchi}
\partial_\mu W_s +W_s \partial_\mu \si = -
\na_\nu T_m{}^\nu_{\mu}.
\ee
 This  is a  consistency condition for $\si$ 
 (supplementing its FE~(\ref{FEs})) 
 being,  with account for (\ref{W}), of the third order 
 what,  in a general case, may encounter some instabilities.

 \subsection{Special case}
 
\subsubsection{Matter conservation}

The general solution to (\ref{Bianchi}) may be looked-for 
in the form $W_s =W_0(x) e^{-\si}$, with $\partial_\mu W_0 = -
e^\si \na_\nu T_m{}^\nu_{\mu}$.
A crucial   simplification occurs in the case if  $L_m$ is independent of $\si$, 
$\de L_m/\de\si=0$,
so that $T_{m\mu\nu}$ is covariantly 
conserved/continuous per se, $ \na_\nu T_m{}^\nu_{\mu}=0$.\footnote{This
may  in reality be an oversimplification,  with a direct correlation of the scalar graviton 
and matter becoming, conceivably, in some cases even  crucial, 
e.g., in the case of the  scalar-graviton DM  halos of  galaxies
to match with  the well-known Tully-Fisher law.}
In this case (or under  the  absence of matter),  
(\ref{Bianchi}) possesses the  first integral  
(playing the role  a global degree of freedom) reducing 
the order of the equation for $W_s$ to the second one:
\be\label{W_s}
W_s = \ka_s^2 \na^\la\na_\la \si +V'_s   =\ka_g^2 \La_0 e^{-\si}, 
\ee
 where  there  is put $W_0= \ka_g^2 \La_0$, with   $\La_0$ 
an  integration constant.
With  $U_s=V_s+W_s$ becoming now   the  bona fide  
effective scalar potential $U_s\equiv V_s+\ka_g^2 \La_0 e^{-\si}$,
the scalar field  satisfies  the canonical
second-order FE: 
\be\label{W0}
\ka_s^2  \na^\la\na_\la \si +U'_s=0,
\ee
At that, 
the Ostrogradsky instability potentially possible for a solution of the third-order
eq.~(\ref{Bianchi})  is explicitly eliminated. 

At last, accounting  for (\ref{W_s}) and  (\ref{si}) one can  in this case  present 
FE (\ref{FEs}) for ${\om}^{a}$ at
$W_0\neq 0$   as follows:
\be\label{Qa}
\g  \na_\la(e^{-\si}{\om}^{-1}{}^\la_{a}) =   \partial_\la( \g
e^{-\si}{\om}^{-1}{}^\la_{a})
 =  \partial_\la( \sqrt{-{ \om}}\,
{\om}^{-1}{}^\la_{a})=0,
\ee
which  proves   to be independent of $g_{\mu\nu}$. 
To put it differently, this FE expresses 
the  conservation of the GC current density 
${\cal J}^\la_a=\sqrt{-\om}{\om}^{-1}{}^\la_{a}$. 
Now, having found from FEs
the metric and $\si$,  and extracting
hereof  ${\om}=g e^{-2\si}$, one can  find
the proper ${\om}^{a}$  up to a residual freedom
consistent with  the required ${\om}$.
Such an ambiguity   is insignificant and 
may, in principle, be eliminated   afterwards in a more complete  theory.

\subsubsection{Scalar graviton: dark matter vs.\ dark energy}

Altogether, depending on $\La_0$ 
there are conceivable three generic cases with the principally different behavior 
for the scalar graviton as an emergent dark substance.\footnote{Note that 
in the case of   $L_m$ being dependent on $\si$
the value and sign  of $\La_0$ 
 for the different solutions  (treated   
in the different spacetime regions as approximations to an exact one) 
may be allowed to vary, in distinction
with the parameters of $V_s$ fixed  ab initio for all the solutions.}

({\em i}) $\La_0<0$ arbitrary (varying). This case may be argued to   
be associated with  the stationary scalar-graviton field in the closed spatial regions 
corresponding to the galaxy DM halos.\footnote{In favor of such a possibility
there was  argued in~\cite{Pir3}, though in the frameworks with a 
non-dynamical auxiliary scalar density $\om$. For a dynamical $\om$,
this case in the context  of DM remains to be investigated.}

({\em ii}) $\La_0=0$. This intermediate case corresponds to the 
scalar graviton as a canonical scalar field.\footnote{For a  
canonical (ultra-)light scalar field  in the context 
of  the galaxy DM halos, see, e.g.~\cite{ Hu, Lee}.}  

({\em iii}) $\La_0>0$ arbitrary (fixed). 
We associate  this  case below  
with  the homogeneous  scalar-graviton field  as DE
filling-up the Universe as a whole.\footnote{It may 
thus be said  that the term 
$W_s$ in the effective scalar potential, being completely  ad hoc 
for a canonical scalar field and drastically influencing the  manifestations
of the latter, is  a kind of the ``Black Swan'' for the scalar graviton. 
The appearance of $W_s$ may, in turn, be traced back to the 
combined (dependent, in particular,  on  metric) nature of the scalar graviton, 
$\si=\si(g/{\om})$.}  

In what follows, we concentrate on the 
scalar-graviton DE alone and  do not dwell 
into the specific nature of  DM, in particular, is it (and how) 
associated with the scalar graviton or not.

\section{Scalar graviton as dark energy}
\label{sec:4}

\subsection{General case}

Now we apply the   results  above  to the Universe as a whole.
Assuming the latter to be homogeneous and isotropic
choose conventionally the FRW metric   given  
in the standard cosmological coordinates $x^\mu=(t, r, \theta,
\varphi)$  by the line element
\be
d s^2= d t^2- a^2\Big (\frac{1}{1-K r^2} d r^2 +r^2 d\Omega^2\Big).
\ee
Here $t$ is the  standard cosmological  time, $r $  the  radial
distance from an (arbitrary chosen) spatial origin,
$a(t)$ a scale factor, $K=k/R_0^2$, with $R_0$ an arbitrary fixed  unit of
length,  and $k=0,\pm 1$ for the zero, positive and negative spatial curvature
of the spatially flat, closed and open Universe, 
respectively.  Let  the Universe  be filled-up with  a
continuous medium/matter (taken for simplicity to be of one kind)  
possessing  the energy-momentum tensor
\be
T_m^{\mu\nu} =(\rho_m+p_m) u^\mu u^\nu - p_m g^{\mu\nu},
\ee
where   ${\rho}_m(t)$ and $p_m(t)$ are,  respectively,  the medium energy density
and pressure,  and $u^\mu$ ($u^\la u_\la =1$)  the medium comoving four-velocity, with
$u^\mu=(1,0,0,0)$ in the standard cosmological coordinates. 
The same, with $\rho_{DM}$ and $p_{DM}$, is assumed for a conceivable DM.
Additionally, these substances are assumed to be characterized  
by some,  given  ab initio, indices of state, 
$w_m=\rho_m/p_m$ and $w_{DM}=\rho_{DM}/p_{DM}$, respectively ($w_{DM}=0$ 
for a cold DM (CDM)).
In the  spirit of $\La$CDM,  the total
energy density and pressure of the Universe, ${\rho}$ and $p$, 
in the presence of the scalar graviton $s$
are  given by the sum of the four  fractions:
\bea\label{tot}
 {\rho} &=&  {\rho}_m+   \rho_{DM}+\rho_\La + 
 \rho_{s}\equiv  {\rho}_M+  {\rho} _s ^\La\nn\\
 p&= &p_m+ p_{DM} +p_\La + p_{s} \equiv  p_M+p_s^\La,
\eea 
with $M=(m, DM)$ referring to  the total matter, incorporating the ordinary one
and DM, and the the Lagrangian CC $\La$ (with $w_\La=-1$). The latter will, 
in a general case,  be  included  into a constant part of a 
redefined scalar-graviton  potential, with  the superscript $\La$ 
being in what follows omitted.
In these terms, the Friedman-Lema\^itre  gravity equations  
for the evolution of the Universe   look
like:
\bea\label{FL}
\ddot a/a &=& - ({\rho}+3p) / 6\ka_g^2  ,\nn\\
H^2+K/a^2&=&   {\rho}/3 \ka_g^2,
\eea
with  a dot meaning a time
derivative,  and the Hubble parameter $H\equiv \dot a/a$   giving the 
relative expansion rate of the Universe. 

At that, the homogeneous 
scalar-graviton  field,
$\si(t)$, is treated as the  omnipresent DE  spilled all over the Universe.
More particularly, one gets\footnote{The role of 
$u^\mu$ here plays  $n^\mu= \na^\mu \si/( \na^\la \si  \na_\la
\si)^{1/2}$, which at $\si=\si(t)$ is  $n^\mu =(1,0,0,0)$ in the
standard cosmological coordinates.}
\bea\label{DE}
{\rho}_{s}&=&\frac{1}{2}\ka_s^2  \dot \si^2  + U_s ,\nn\\
 p_{s} &=&\frac{1}{2}\ka_s^2  \dot \si^2  - U_s.
\eea
In the  above,  $U_s=V_s+W_s$ is the effective scalar-graviton potential, with $V_s$ the 
Lagrangian scalar potential,
and the scalar  wave operator $W_s$ as follows:
\be\label{Ws}
W_s\equiv -\de L_s/\de \si =\ka_s^2(\ddot\si+3H  
\dot \si)+  \partial V_s/\partial \si .
\ee
These expressions are valid at any  $k$ 
and  correspond to the scalar-graviton DE
with the variable   effective index  of state $w_{s}(\si)\equiv p_s/\rho_s$.
Under  weakly changing $\si$, $\dot \si\simeq0$, (though, generally, $\ddot
\si\neq 0$) one has $w_{s}=-1$
mimicking thus the $\La$-term. 
The second time-derivative of $\si$ generally appears in ${\rho}_{s}$ and  $p_{s}$, 
even under  the simplest Lagrangian $L_{sg}$,
through the off-shell contribution $W_s$ due to the intrinsic dependence of $\si$ 
on metric.
  
The evolution equations (\ref{FL}) are to be 
supplemented by the  covariant conservation/continuity condition  for the scalar graviton  
$s$ and  the total  matter~$M$:
\be\label{WM}
  \dot W_s  + W_s\dot \si= -(  \dot {\rho}_{M}+3 H({\rho}_{M}+ p_{M})),
\ee
which   follows from the reduced Bianchi  identify. 
Generally,  this is the third-order equation for $\si$ due to
a correlation of  the scalar-graviton field as  DE and  the total matter. 

The equations above
derivable in the multi-component gravity  under the quadratic-scalar reduction
describe  the  looked-for 
scenario for  the evolution of the homogeneous
isotropic Universe filled-up with the ordinary matter, DM and the scalar gravitons.  
Having
found hereof  $a(t)$ and $\si(t)$, and  
thus (at $k=0$)  $\sqrt{-g}=a^3$ and  $\sqrt{- \om}=e^{-\si} a^3$,
one can  then get  
from (\ref{Qa})  the inverse tetrad $\om^{-1}{}^\la_a$:
\be\label{tetrads}
{\om}^{-1}{}_{0}^0\sim 1/\sqrt{-{\om}}, \ \  {\om}^{-1}{}_{\al}^l = \de_{\al}^l,
\ee
where $a=(0,\al)$, $\la=(0, l)$; $\al, l =1,2,3$.  
Inverting 
(\ref{tetrads}), so that $\om^0_0\equiv\partial_0\om ^0 \sim \sqrt{-\om}$ and 
 ${\om}{}^{\al}_l   \equiv \partial_l \om^\al= \de^{\al}_l$,     
and choosing properly the coefficients one gets
\be\label{qat}
{\om}^{0}=  \int \om^0_0 dt+c^0=  \int  \sqrt{-{ \om}}d t +c^0,\ \ 
{\om}^{\al}=\de^{\al}_l x^l +c^\al
\ee 
defined  up to  some integration constants $c^0$  and $c^\al$ (to be put 
for simplicity zero), and 
$\det(\partial_\la
{\om}^{a})=\sqrt{-{\om}}$, as it should be.
As for the quasi-affine coordinates 
$\mathring{x}^a=(\mathring{t}, \mathring{x}^\al)$,   eq.~(\ref{qat}) determines 
the  quasi-affine time $\mathring{t}\equiv {\om}^{0}(t)$, 
related with the cosmological one through
$d \mathring{t}= \sqrt{-\om}\, d t$, and the spatial quasi-affine coordinates 
$\mathring{x}^\al$, coinciding (at $k=0$) with the cosmological ones.

\subsection{Special case}

\subsubsection{Matter conservation}

A  significant simplification of the preceding consideration occurs
if each of the matter components $M= (m, DM)$ is independent of $\si$, and thus 
covariantly conserved/continu\-ous, with the r.h.s.\ of 
(\ref{WM})  being zero\footnote{Though, this may, generally, be 
an oversimplification, especially what concerns DM.}
\be
\dot {\rho}_{M}+3 H({\rho}_{M}+ p_{M})=0.
\ee
Then it follows  as before that $W_s=\ka_g^2 \La_0 e^{-\si}$,
with $\La_0$ an integration  constant taken to be positive. 
eq.~(\ref{Ws}) then implies, in turn,  that  the scalar-graviton FE is 
determined at any $k$ by the effective potential $U_s =V_s+\ka_g^2 \La_0
e^{-\si}$~as
\be\label{U_s}
\ddot \si+3 H  \dot \si+\ka_s^{-2}\partial U_s / \partial \si =0.
\ee
Let now $\bar\si$ be the position of the minimum of the effective potential,
$\partial U_s/\partial \si|_{\bar\si}=0$.
Under neglecting   by $\dot\si$ and  $\ddot\si$  this  FE   reduces to
$\partial U_s / \partial \si =0$, meaning  $\si$  to be  restricted by
$\bar\si$. 
By this token,  designating
$U_s|_{\bar\si} \equiv \ka_g^2 \bar \La_s$ and  replacing in
${\rho}_{s}$ and $p_{s}$ the $\si$-dependent $U_s$ by the constant
$U_s|_{\bar\si}$, 
one arrives  (assuming  $w_{DM}=0$) at the standard $\La$CDM
model,  corresponding to  the  effective  CC  
$\bar \La_s$  with
\be
\bar{\rho}_{s}=- \bar p_{s}=\ka_g^2\bar\La_s, 
\ee
reproducing thus in such an approximation  the standard $\La$CDM model.

\subsubsection{Scalar-graviton dominance}

To proceed explicitly further, consider the evolution of
the Universe after a long time of its preceding expansion.  
Adopt at such a   late-time stage  
 the overwhelming dominance  of DE by putting\footnote{Qualitatively,
 this may be not unrealistic due to the ratio of the energy 
 of  the total matter ($M =(m, DM)$)   
 to DE at the present epoch being $0.3:0.7$.}
\be
{\rho}_M=p_M=0. 
\ee
With account for  the second part of FEs~(\ref{FL}),  eq.~(\ref{U_s}) 
acquires  at $k=0$ an autonomous form as follows:
\be\label{PS}
\ddot \si+\sqrt{3}   \upsilon_s  \Big( \frac{1}{2} \dot \si^2
+ U_s/\ka_s^{2}  \Big)^{1/2} \dot \si+\ka_s^{-2}\partial U_s /
\partial \si =0,
\ee
where 
\be
\upsilon_s\equiv \ka_s/\ka_g
\ee
(supposedly, $\upsilon_s \ll 1$).
This is the  master equation for
the  evolution of the  Universe due to the pure scalar-graviton DE.
In particular, $\si\equiv\bar \si$ is the exact solution to the equation. 
Having found from (\ref{PS})  $\si$ one  can  then find from 
the second part of the  Friedman-Lema\^itre equations (\ref{FL}) at $k=0$
the Hubble parameter 
\be
H\equiv\dot a/a={\rho}_{s}^{1/2}/
\sqrt{3}\ka_g \ge 0. 
\ee
and the respective scale factor
\be\label{a} 
a= a_0 \exp \frac{1}{\sqrt{3}\ka_g}\int  \rho_s^{1/2}dt= 
a_0 \exp\frac{ \upsilon_s }{\sqrt{3}}  \int \Big( 
\frac{1}{2}\dot
\si^2 + U_s/ \ka_s^2 \Big )^{1/2}d t  , 
\ee
with $a_0$ an integration constant,  
To envisage  the behavior
of $H(t)$  note that
combining (\ref{FL}) at $k=0$ one can find 
\be\label{Hsi}
\dot H =  -\frac{1}{2}(\rho_s +p_s)= - \frac{1}{2}\upsilon_s^2 \dot \si^2\leq 0, 
\ee 
independently of $U_s$. This means, in particular, that  $H$  always
monotonically decays approaching  atop  a constant
value~$\bar H\ge 0$.
At that, the Hubble horizon   $H^{-1}$   monotonically expands to $\bar H^{-1}$.

\subsubsection{Effective cosmological constant}

More specifically, let us put for the Lagrangian CC $\La=0$ and
choose the scalar potential  quadratic in $\si$:
\be 
V_s =\frac{1}{2}m_s^2 (\ka_s \si)^2\equiv \frac{1}{2}\ka_g^2 \mu_s^2\si^2,
\ee
 with $m_s$ the scalar-graviton 
mass and $\mu_s\equiv \upsilon_s m_s$ its reduced mass, so that
\be\label{CW}
U_s=
\ka_g^2\Big(\frac{1}{2}\mu_s^2 \si^2+\La_0 e^{-\si}\Big),
\ee
Rewrite the  scalar-graviton  FE in this case  as follows:
\be\label{PS'}
\si''+\sqrt{3}  \upsilon_s   \Big( \frac{1}{2}  \si'^2
+ \frac{1}{2} \si^2 +\al e^{-\si} \Big)^{1/2} \si'+   \si -\al e^{-\si} =0,
\ee
where  
\be
\label{al}
\al\equiv \La_0/\mu_s^2=\La_0/ \upsilon_s^2 m_s^2
\ee
and  $\si'\equiv d \si/d \tau$, with 
$\tau =m_s t$.
In these terms, the expansion rate at  $k=0$ reads 
\be\label{H/m}
H/m_s=a'/a=  (\upsilon_s/\sqrt{3}) 
\Big(\si'^2/2+ \si^2/2+\al e^{-\si}\Big)^{1/2}.
\ee

The evolution of the Universe 
entirely due the scalar-graviton DE is given by solutions 
to (\ref{PS'}) and (\ref{H/m}) 
presented  below. 
The behavior of the looked-for solution is mainly determined  by  
the minimum of the effective potential $U_s(\si)$ eq.~(\ref{CW}).
For  the position $\bar \si$ of such a   minimum 
and the respective effective CC 
$ \bar\La_s\equiv U_s|_{\bar \si}/\ka_g^2=\mu_s^2(\bar\si^2/2+\al
e^{-\bar\si})$  as a function of 
$0\leq \La_0,<\infty$  at a given  $\mu_s$  
  one roughly encounters   the following generic cases:
\bea\label{cases}
(a)   & \al \ll 1,  & \bar \si \simeq \al,  \ \ \ \    \bar\La_s \simeq  \al 
\mu_s^2 =  \La_0,   \nn\\
(b)  & \al\sim 1, & \bar \si \sim 1,  \  \  \ \ \, 
\bar\La_s \sim \mu_s^2\sim \La_0,  \nn \\
(c)  & \al    \gg 1, & \bar \si \simeq\ln \al, \ \,
 \bar\La_s\simeq  \Big( (1/2) 
\ln^2 \al+1\Big)\mu_s^2.
\eea
According to (\ref{al}) the case (a)  corresponds  to the relatively large
Lagrangian potential $V_s$ compared to $W_s$,
the  case (c) to the relatively large spontaneous contribution $W_s$ compared to $V_s$,
and the  case (b) to the approximate equality of both contributions.
Otherwise, fixing $\La_0$
one sees that 
the  three generic cases (a)--(c) in ~(\ref{cases})
refer, respectively,   to the scalar graviton relatively heavy ($\mu_s\gg\sqrt{\La_0}$),
intermediate  ($\mu_s\sim \sqrt{\La_0}$) 
and (ultra-)light  ($\mu_s\ll\sqrt{\La_0}$). In the last case,  due to 
$\bar \La_s\sim \upsilon_s^2 m_s^2$  the limit $m_s\to 0$  or $\upsilon_s\to 0$  
naturally, in a technical sense.  assures the limit  $\bar \La_s\to 0$.

The behavior of the respective solution to FEs on the phase
plot $(\si,\si')$ for the case $\al=1$,  starting from a freely chosen initial point
$(\si_0=- 5, \si'_0=0)$, is shown in  figure~1.  
It is seen  that  the solution tends to a point  $(\bar\si, 0)$ serving 
as  an  attractor. The solution winds  clock-wise 
around  the attractor approaching the latter asymptotically at  $\tau\to\infty$. 
A similar behavior can be shown to tale place at other values 
of $\al$ considered previously. The value $\upsilon_s=0.1$ is chosen 
just for  illustration purposes. 
For the smaller $\upsilon_s$ the tightness of winding increases (under unchanged~$\al$),
with the picture becoming less prominent. 
The behavior of $H/m_s$ vs.\ $\tau=m_s t$   extracted numerically 
from  figure~1 and the similar ones for other $\al$, with  the initial time $\tau_0=1$  
attributed to the initial point $(\si_0=-0.5; 0)$, is shown in figure~2.  
As an independent  test, eq.~(\ref{Hsi}) rewritten as 
\be
H'/m_s\upsilon_s^2 \si'^2= -1/2
\ee
proves to be numerically  valid up to a proper accuracy for  the  numerical 
$\si(\tau)$ and $H(\tau)/m_s$.  
Clearly, $H$ approaches monotonically 
the asymptotic value  $\bar H$ at $\tau=m_s t\to \infty$. 
Imposing for definiteness the asymptotic value of the Hubble parameter 
$\bar H=75\,  km/s /Mpc =   1/4\,  c/Gpc$, 
one can  infer from figure~2 
for the respective Compton wave lengths of the 
scalar graviton  (in the units $ c=\hbar=1$)  the following 
values:\footnote{For comparison, $m_s=1/Gpc  \simeq 10^{-28}eV/c^2 $ 
in the conventional  units.} 
\bea
(a1) &\al& =0.01, \ \, 1/m_s\simeq 3 \cdot 10^{-2} Gpc ; \nn\\
(a2)  &\al &=0.1, \ \ \ 1/m_s\simeq 10^{-1}  Gpc ;\nn\\ 
(b)  & \al &=1,  \ \ \ \ \ \, 1/m_s\simeq 3 \cdot 10^{-1}  Gpc   ;\nn\\
(c) &\al &=100, \ \ \,   1/m_s \simeq 1Gpc .
\eea
Stress that such values of the Compton wave length, less or of the order 
of the Universe size, are just the representative ones 
corresponding to  the (rather loosely)  chosen values of  the parameters
$\upsilon_s$ and $\al$.

It  follows from~(\ref{a})  that at $k=0$  the attractor 
produces  the exponential expansion
\be\label{scalfact}
  \bar a(t) = a_0 \exp (\bar\La_s/3)^{1/2} t \equiv a_0 \exp \bar H t.  
  \ee 
In other words, there  takes place 
 the late-time   inflation corresponding to the spontaneously emerging
 at the level of FEs  the effective CC  $\bar \La_s$ even
 under the absence of the  Lagrangian  CC.\footnote{This may to an extent  
be  reminiscent of the 
Unimodular Relativity(UR)/Unimodular Gravity/TDiff, 
missing the  scalar graviton   effectively
due to the  restriction $g=\om$, with a non-dynamical $\om$. 
Here  CC is also not  a Lagrangian parameter $\La$ 
but an integration constant~$\La_0$ appearing
spontaneously at the level of~FEs. For  the (modified) 
UR, cf., e.g.,~\cite{Saez}, with the numerous 
references therein.}
Had any of $\La_0$ or $\mu_s$ been
zero the asymptotic inflation would not take place (under the assumed $V_s|_{\rm
min}=\ka_g^2\La=0$).
By this token,   having adopted   the vanishing of the Lagrangian  CC $\La$, 
one could eventually explain 
the effective CC~$\bar\La_s$ to be tiny but  nonzero, showing the way  
to partially solving one of the CC problems. Still, justifying   the 
Lagrangian CC being zero and ensuring the quantum stability 
of the  tiny classical effective CC is beyond the scope 
of the present paper.\footnote{For the CC problems, see, e.g.~\cite{Wein},
and for some recent (far from being exhaustive) attempts at 
solving them, cf., e.g., \cite{Carb}--\cite{Oda} and \cite{Charm1}--\cite{Noj}.} 
  
  As a final remark, at $\si=\bar \si$ due to  
$\sqrt{-\bar \om}\sim   \sqrt{-\bar g}= \bar a^3\sim \exp 3\bar H t$,  the  
quasi-affine  time $\mathring{t}$ as defined  
by  the attractor  at $k=0$   is 
$\mathring{t}= t_0 \exp 3\bar H t$, or
inversely $t= (3\bar H)^{-1} \ln \mathring{t}/ t_0$, with  $t_0$ an
integration constant.  eq.~(\ref{scalfact}) implies 
then that $\bar a=a_0 (\mathring{t}/t_0)^{1/3}$,  with   
the  characteristic volume ($\bar
a^3= a_0^3\, \mathring{t}/ t_0$)   and the characteristic energy
($\ka_g^2\bar \La_s \bar a^3$) of the
Universe increasing  linearly in  the quasi-affine  time~$\mathring{t}$ 
(or, rather, v.v.), with the late-time 
inflation appearing just  in disguise.  
Conceivably, this may  present an alternative view on   the evolution of the Universe 
and the meaning of time.

\section{Conclusion}
\label{sec:5}
 
In conclusion,  the scalar-graviton reduction of   the multi-component gravity, 
with  the massless tensor graviton supplemented by the massive scalar one,
may present the viable modified gravity nearest-to-GR.
At that, the scalar graviton proves to be quite plausible candidate on the role of 
a dark substance of the Universe: DM and/or DE depending on the solution.
The scalar graviton as DE  may,  
in a natural way,  explain 
the appearance at the classical level  of a nonzero but tiny effective CC 
due to the attractor solution for the (ultra-)light scalar graviton. 
Such a signature of the scalar-graviton DE makes  going beyond the 
$\La$CDM model per scalar graviton quite promising.
Further studying  the scalar graviton
to validate it as an emergent dark substance of the Universe is urgent.

\paragraph{Acknowledgment}
The author is  grateful to  I.Y.~Polev for assistance with  graphics.

\newpage

\begin{figure}
\begin{center}
\resizebox{0.5 \textwidth}{!}{%
\includegraphics{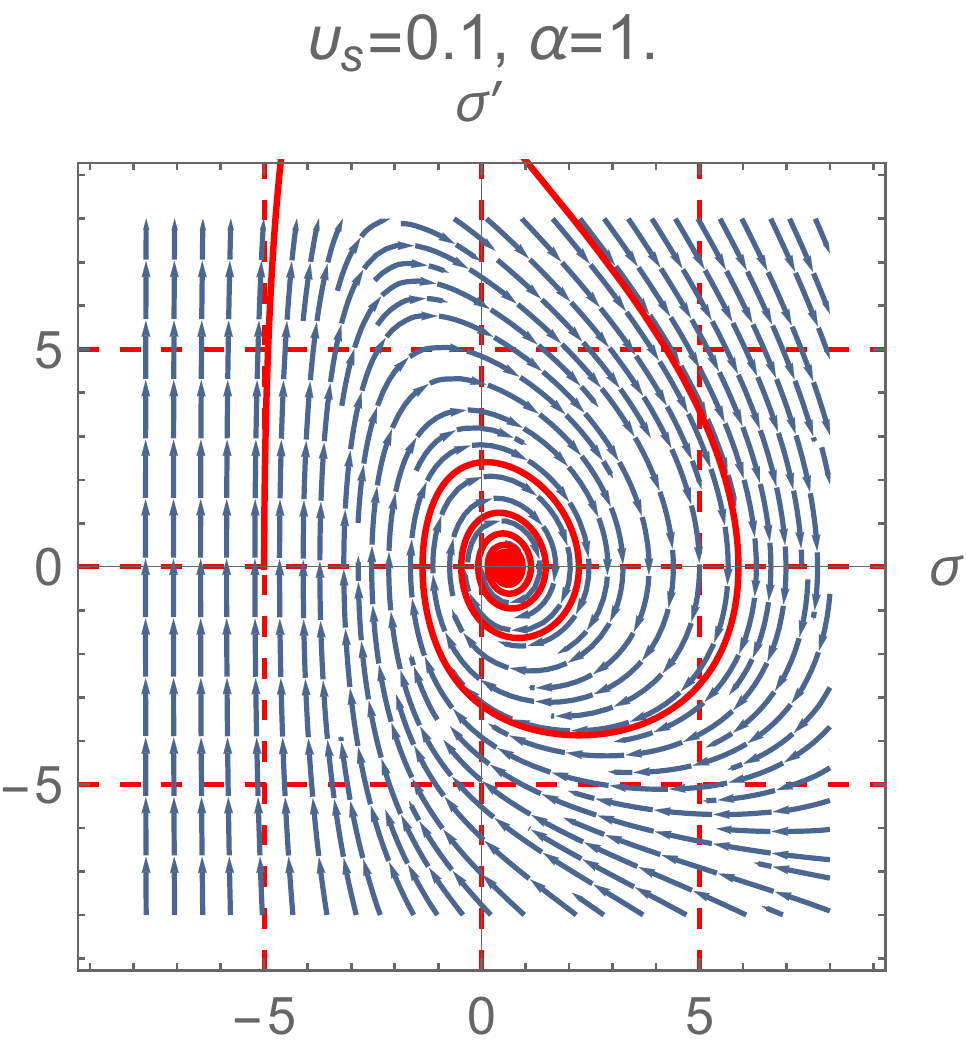}}
\end{center}
\caption{The phase plot $(\si,\si')$ 
describing  the  (clock-wise) evolution  of the scalar-graviton 
DE  (thick line) depending on $\tau=m_s t$ 
at the representative values of parameters 
$\upsilon_s=\ka_s/\ka_g=0.1$ 
and $\al \equiv \La_0/\upsilon_s^2 m_s^2=1$. 
The thin lines are the flow ones.
The initial point $(-5, 0)$ corresponds to the initial time $\tau_0=1$.
Independent of the initial point, all the solutions approach at $\tau\to \infty$ 
the same  attractor point $(\bar \si,0)$  corresponding  
to an emergent effective CC $\bar \La_s$. 
A similar behavior tales place for other values of $\al$.
\label{fig:1}}
\end{figure}
\begin{figure}
\begin{center}
\resizebox{0.6 \textwidth}{!}{%
\includegraphics{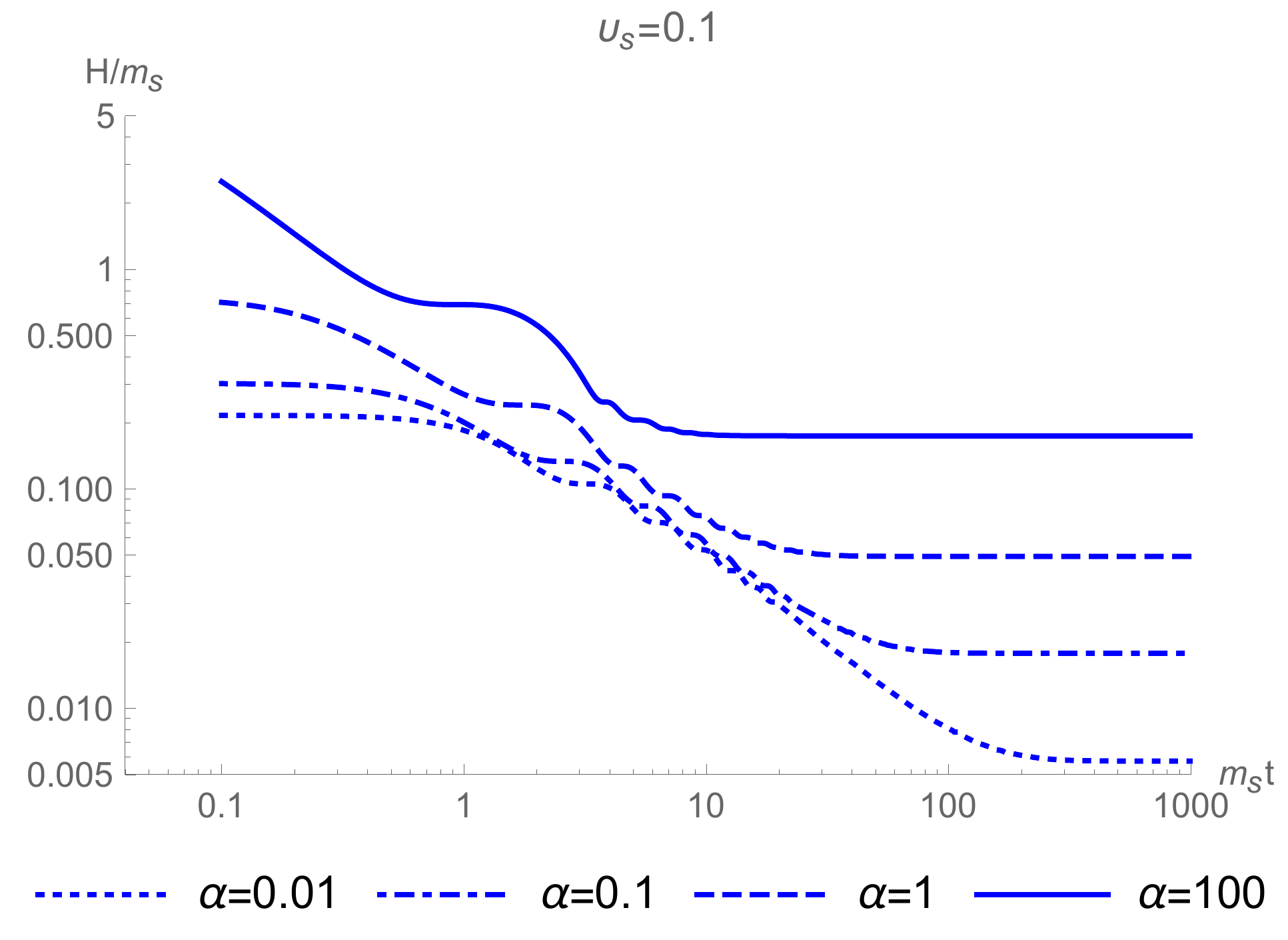}}
\end{center}
\caption{The normalized Hubble scale $H/m_s$ vs.\  $\tau=m_s t$
for the representative cases (a)--(c) in terms of  the 
scalar-graviton mass $m_s$ (in the units $\hbar =c=1$). 
With the asymptotic Hubble constant
fixed at $\bar H=\sqrt{\bar \La_s/3}=75 km/s/Mpc$, 
the  implied Compton wave lengths  
$\la_s/2\pi=1/m_s $ of the scalar graviton are:  
(a1) $\al =0.01$, $1/m_s\simeq 30 Mpc$;  
(a2) $\al =0.1$, $1/m_s\simeq 10^2 Mpc$; 
(b) $\al =1$, $1/m_s\simeq3\cdot 10^2 Mpc$; 
(c) $\al =100$, $1/m_s\simeq 10^3 Mpc$. 
(For comparison,  $m_s=1/ Mpc \simeq 10^{-25} eV/c^2$ 
in the conventional units,)
\label{fig:2}}
\end{figure}

\end{document}